\documentclass[final,5p,times,twocolumn]{elsarticle}

\usepackage{amssymb}
\usepackage{multirow}
\usepackage{appendix}
\usepackage{verbatim}
\usepackage{mathtools}
\usepackage{float}
\usepackage{hyperref}
\usepackage{balance}
\usepackage{amsthm}
\usepackage{amsmath}
\usepackage{comment}
\usepackage{caption}
\usepackage{graphicx}
\usepackage{subcaption}
\usepackage{threeparttable}
\usepackage{epstopdf}
\usepackage{color}     
\usepackage{array}
\newcommand{\tabincell}[2]{\begin{tabular}{@{}#1@{}}#2\end{tabular}}
\usepackage{longtable}

\journal{}

\begin{document}

\begin{frontmatter}



\title{Discovery potential for supernova relic neutrinos with slow liquid scintillator detectors}

\author[a,b]{Hanyu Wei\corref{why}}
\ead{hwei@bnl.gov}
\author[a,b]{Zhe Wang}
\author[a,b]{Shaomin Chen}
\cortext[why]{Corresponding author.}
\address[a]{Key Laboratory of Particle \& Radiation Imaging (Tsinghua University), Ministry of Education, Beijing 100084, China}
\address[b]{Department of Engineering Physics, Tsinghua University, Beijing 100084, China}

\begin{abstract}
Detection of supernova relic neutrinos could provide key support for our current understanding of stellar and cosmological evolution,
and precise measurements of these neutrinos could yield novel insights into the universe.
In this paper, we studied the detection potential of supernova relic neutrinos using linear alkyl benzene (LAB) as a slow liquid scintillator. The linear alkyl benzene features good separation of Cherenkov and scintillation lights, thereby providing a new route for particle identification.
We further addressed key issues in current experiments, including (1) the charged current background of atmospheric neutrinos in water Cherenkov detectors and (2) the neutral current background of atmospheric neutrinos in typical liquid scintillator detectors. 
A kiloton-scale LAB detector at Jinping with $\mathcal{O}$(10) years of data could discover supernova relic neutrinos with a sensitivity comparable to that of large-volume water Cherenkov detectors, typical liquid scintillator detectors, and liquid argon detectors.

\end{abstract}


\end{frontmatter}


\section{Introduction}
Galactic core-collapse supernovae are estimated to occur at a rate of only a few per century~\cite{Rate},
and 99\% of the gravitational binding energy of these events is carried away by neutrinos.
Besides detection of neutrino bursts from explosion of a supernova, such as 1987A~\cite{Hirata1,Hirata2,Bionta,Bratton,Alekseev1,Alekseev2}, detection of supernova relic neutrinos, SRNs, also known as the diffuse supernova neutrino background, DSNB, can also be expected. These neutrinos originate from an enormous number of supernova explosions throughout the time and space of the universe and can provide researchers with novel insights into stellar evolution and cosmology.

Current experimental upper limits were obtained from the SNO experiment in 2006~\cite{SNO2006}, the KamLAND experiment in 2012~\cite{Kam2012}, and the Super-Kamiokande (SK) experiments in 2012~\cite{SK2012} and 2015~\cite{SKnH}.
Relatively low signal-to-background ratios are the main problem of these studies.
In the future, gadolinium-doped water detectors~\cite{Gdwater, SKntag, SKGdstatus}, typical liquid scintillator detectors with the scintillation light pulse shape discrimination~\cite{JUNO, JUNOphysics, LENA, LENA-SRN}, or liquid argon time projection chamber detectors~\cite{LArFirst, LArUS, DUNE} may come online to improve findings in this filed of research. 

Slow liquid scintillators feature scintillation emission time significantly longer than the Cherenkov emission time and may thus help identify different particles. Linear alkyl benzene (LAB) has been suggested to be a candidate target material in slow liquid scintillators for the future Jinping neutrino experiment~\cite{Jinping} because it can be directly used or improved to increase light yield by mixing wave-length shifters. 
Using the results measured in Ref.~\cite{LAB}, we studied the potential of using LAB as a realistic candidate target material to detect SRN.

This paper is organized as follows: We summarize the key issues and possible solutions of SRN detection in Section~\ref{sec:KeyIssues}, present the ability of LAB in particle identification (PID) in Section~\ref{sec:slowscint}, report the sensitivity of SRN detection with LAB at Jinping in Section~\ref{sec:sens}, and then conclude our study in Section~\ref{sec:conclusion}.


\section{SRN detection}
\label{sec:KeyIssues}
\subsection{SRN signal}
\label{sec:srnmodel}
The differential SRN flux, $d\phi(E)/dE$, is calculated by integrating the rate of core-collapse supernovae, $R_{\rm ccSN}$($z$), the energy spectrum of neutrino emission, $dN/dE$, and the redshift, $z$, over the cosmic time,~\cite{SRNmodel} 
\begin{equation}
\label{equ:srn}
\frac{d\phi(E)}{dE}=c\int{R_{\rm ccSN}(z) \frac{dN(E')}{dE'}}(1+z)\left|\frac{dt}{dz}\right|dz
\end{equation}
where $\left| dz/dt \right| = H_0(1+z)[\Omega_m(1+z)^{3}+\Omega_{\Lambda}]^{1/2}$ and $E'=E(1+z)$. $H_0, \Omega_m$, and $\Omega_{\Lambda}$ are cosmological parameters.

Many SRN models haven been constructed to predict both the flux and the spectrum. Earlier models were developed before SN1987A~\cite{Model1, Model2, Model3}, and even more sophisticated models~\cite{ConstSN, CosmicGas, ChemicalEvo, HeavyMetal, LMA, FailedSN, HBD} were proposed after SN1987A.
The shapes of the predicted SRN fluxes are similar among these models. In this paper, we adopted a recent HBD model described in Ref.~\cite{HBD} by Horiuchi, Beacom and Dwek. The effective neutrino temperature, 6 MeV, in the HBD model is used in this paper as a typical value.

\subsection{Detection in hydrogen-rich detector}
In a hydrogen-rich detector, supernova relic neutrinos and supernova burst neutrinos are primarily detected via inverse beta decay, IBD, $\bar{\nu}_{e}+p\rightarrow n+e^{+}$, due to their large cross section, which is about 2 orders of magnitude greater than the next most-frequent interaction channels~\cite{SNdetection}, for instance, elastic scattering on electrons and charged/neutral current scattering on entire nuclei.
Notice that in a liquid argon time projection chamber detector, LArTPC, SRN events are detected via the neutrino charged and neutral current interactions with argon nuclei as well as elastic scattering on argon electrons. In the case where energies are lower than $\sim$20 MeV, the dominant backgrounds for LArTPC in the SRN study come from $^8$B and \emph{hep} solar neutrinos~\cite{LArSRN}. Since LArTPC presents a fairly unique detection technique and background, as well as the heavy water Cherenkov detector such as SNO~\cite{SNO2006}, this section focuses on the hydrogen-rich detector.

\subsubsection{Detection techniques}
\emph{Liquid scintillator (LS)}: A prompt-delayed coincident measurement of an IBD event is generally performed in LS, based on the scintillation photons emitted once a charged particle deposits energy in it. The prompt signal is from the deposited energy of the positron and its annihilation gammas. The delayed emission of gamma(s) is given by the neutron capture on hydrogen or doped isotopes (e.g., gadolinium (Gd)).
The coincidence from the prompt and delayed signals provides a clear signature against the backgrounds from the accidentals, radioactivity, and other flavors of neutrinos with different interactions, for instance, solar neutrinos. 

\emph{Water}: A water Cherenkov detector identifies IBD events based on the Cherenkov photons radiated by the IBD positrons. In general, no prompt-delayed coincidence measurement is performed in the water Cherenkov detector, as the 2.2 MeV gamma obtained from neutron capture on hydrogen is difficult to detect (e.g., in the early stages of the SK experiment). A forced trigger could be implemented for the later stage of the SK data to search for a delayed coincident 2.2 MeV signal of neutron capture on hydrogen~\cite{SKnH, SKntag}. The neutron-tagging technique in water allows the powerful coincident measurement of an IBD event based on Cherenkov photons despite a low tagging efficiency.

\emph{Water doped with gadolinium (Gd-water)}: For water doped with Gd, the total energy of the emitted gammas from neutron capture on Gd is $\sim$8 MeV, which enables distinct neutron tagging in comparison with the 2.2 MeV gamma from neutron capture on hydrogen. A high neutron tagging efficiency of about 90\% can be obtained with a 0.2\% of GdCl$_{\rm 3}$-water solution~\cite{SKntag}.

\subsubsection{Background mechanism}
\label{sec:prebkg}

The key backgrounds for SRN detection in current hydrogen-rich detectors are basically categorized into three types: the cosmic-ray muon-induced background, the reactor neutrino background, and the charged and neutral current backgrounds induced by atmospheric neutrinos.

\emph{Cosmic-ray muon-induced backgrounds:}
An energetic cosmic-ray muon interacting with a carbon (oxygen) nucleus can produce a radioactive spallation background that can mimic an SRN signal~\cite{MuonSpall}. This background dominates the SRN analysis in water Cherenkov detectors and can be significantly suppressed by the neutron-tagging technique; an exception to this finding involves the $^9$Li/$^8$He background, which has the exact same signature of an IBD event. The $^9$Li/$^8$He background can also affect the studies in liquid scintillator detectors. The cosmic-ray muon-induced fast neutron is a typical background in liquid scintillator detectors. Neutrons could migrate into the detector and recoil protons or inelastically scatter with carbon nuclei, promptly producing a scintillation signal followed by neutron capture mimicking an IBD event. 

\emph{Reactor neutrino background:}
The $\bar{\nu}_e$'s from nuclear power plants are an indistinguishable background for the SRN search. The energy of reactor neutrinos can be as high as $\sim$10 MeV~\cite{Mueller, Huber}.

\emph{Atmospheric neutrino background:}
The atmospheric neutrino background originates from the four flavors of atmospheric neutrinos, i.e., $\bar{\nu}_e$, $\nu_e$, $\bar{\nu}_{\mu}$, and $\nu_{\mu}$~\cite{Barr1988}.

For charged current interactions with protons or carbon (oxygen) in a detector, 
\begin{itemize}
\item The atmospheric $\bar{\nu}_e$'s form an intrinsic background for the SRN study and are irreducible. Due to the indistinguishable reactor neutrino flux and the atmospheric $\bar{\nu}_e$ flux, a golden window for the SRN study is defined within about 8-30 MeV of neutrino energy~\cite{Gdwater}.
\item The atmospheric $\nu_e$ CC background can be ignored, particularly with the neutron-tagging technique, as the cross sections of various types of atmospheric $\nu_e$ CC interactions with protons or carbon (oxygen) are about two orders of magnitude smaller than the $\bar{\nu}_e$ IBD interaction in the golden window of neutrino energy for the SRN study; the flux is fairly similar to the atmospheric $\bar{\nu}_e$ flux in this energy range. 
\item The atmospheric $\bar{\nu}_{\mu}/\nu_{\mu}$ CC interaction always produces a muon and, in most cases, a neutron even for the $\nu_{\mu}$. The muon can decay into a final state with an electron. These particles would produce Cherenkov lights and probably mimic an IBD event, thereby contaminating the selected IBD sample. The $\bar{\nu}_{\mu}/\nu_{\mu}$ CC interaction has a relatively high energy threshold equal to roughly the mass of a muon (105.7~MeV).   
\end{itemize}

For the neutral current interactions of all the flavors of atmospheric neutrinos, $\pi^{\pm}$ or $\pi^0$ would be produced, and this promptly decays into a final state with a muon or two gammas. An energetic neutron can be produced in some cases, recoiling protons or inelastically scattering with carbon (oxygen) nuclei in the detector. Some isotopes could also be induced with emission of de-excitation gammas. These particles/processes could contaminate the selected IBD sample and produce a main background for the SRN study, particularly in the scintillator detector.

A compilation of different atmospheric neutrino backgrounds and detection techniques is given in Section~\ref{sec:keyissue}.

\subsection{Issues and possible solutions}
\label{sec:keyissue}

The backgrounds induced by cosmic-ray muons and reactor neutrinos are basically crucial for the searches of low-energy SRN events, which relies on the rock overburden and the conditions of the nuclear power plants surrounding the detector. 

The comparative advantages and key issues presented by different SRN detection techniques are presented in Table~\ref{tab:key} in terms of the atmospheric neutrino backgrounds excluding the intrinsic atmospheric $\bar{\nu}_e$ CC background and the ignorable atmospheric $\nu_e$ CC background. 

\begin{table*}[h]
\begin{center}
\begin{threeparttable}
\caption{Comparative advantages and key issues presented by different SRN detection techniques, focusing on atmospheric $\bar{\nu}_{\mu}/\nu_{\mu}$ CC and NC backgrounds. Text in italics reflects the corresponding key issues.}
\label{tab:key}
\begin{tabular}{m{1.8cm}<{\centering} | m{1.4cm}<{\centering} | m{1.6cm}<{\centering} | m{3.2cm}<{\centering} | m{3.0cm}<{\centering} | m{3.0cm}<{\centering}}
\hline
Technique & Signal efficiency & Optical photons & Reactions\tnote{a}& Atmos. $\bar{\nu}_{\mu}/\nu_{\mu}$ CC background & Atmos. NC background  \\
\hline
Water & 75\%~\cite{SK2012} & Cherenkov & \vspace{2mm}  $\bar{\nu}_{\mu}+p\rightarrow \mu^{+}+n$ $\nu_{\mu}/\bar{\nu}_{\mu}+N\rightarrow \mu^{\mp}+({\rm possible:}~\pi^{\pm},~\pi^0)+{\rm anything~else} $ $\nu/\bar{\nu}+N\rightarrow \nu/\bar{\nu}+({\rm possible:}~\pi^{\pm},~\pi^0)+{\rm anything~else}  $\vspace{2mm} & \vspace{2mm} \emph{Difficult to reject decay electrons from invisible $\mu$'s (below Cherenkov threshhold).} \vspace{2mm} & \vspace{2mm} Secondary (decay) products of $\pi^{\pm}, \pi^0, n$ and de-excitation $\gamma$ are mostly ruled out by distinct Cherenkov hit pattern\tnote{b}~ or below Cherenkov threshold. \vspace{2mm}  \\
\hline
Water (Gd-water) with n-tag & \emph{13\%}~\cite{SKnH} (70\%~\cite{SKntag}) & Cherenkov & \multirow{2}{*}{\tabincell{c}{ \\ \\ $\bar{\nu}_{\mu}+p\rightarrow \mu^{+}+n$ \\ $\nu_{\mu}/\bar{\nu}_{\mu}+N\rightarrow$ \\ $\mu^{\mp}+n+{\rm anything~else}$ \\ $\nu/\bar{\nu}+N\rightarrow$ \\ $\nu/\bar{\nu}+n+{\rm anything~else}$} } & \vspace{2mm} With neutron tagging, invisible $\mu$'s are reduced significantly. \emph{ The tagging efficiency is low in water} and increased a lot in Gd-water. \vspace{2mm} & \vspace{2mm}  Further reduced by neutron tagging. \vspace{2mm}   \\
\cline{1-3} \cline{5-6}
Liquid scintillator & 90\%~\cite{Kam2012} & Scintillation &   & \vspace{2mm} The reactions with a $\mu$ below Cherenkov threshold are mostly ruled out by a triple-coincidence of muon, decay electron, and neutron capture. \vspace{2mm} & \vspace{2mm} \emph{ Energetic neutrons from high energy atmospheric neutrinos.} \vspace{2mm}  \\
\hline
\end{tabular}
\begin{tablenotes}
\item[a] $N$ represents a carbon or an oxygen nucleus. In the final state, the nuclide may either be in the ground state or an excited one, probably with one or more neutrons (proton, $\alpha$) scattered off~\cite{Kam2012, LENA-SRN}.
\item[b] Cherenkov hit pattern: Cherenkov angle, number of Cherenkov rings, etc.
\end{tablenotes}
\end{threeparttable}
\end{center}
\end{table*}

\section{Slow liquid scintillator}
\label{sec:slowscint}
Separation of the scintillation and Cherenkov lights in a detector could help reduce the background through PID.

\subsection{Experimental status}
\begin{figure}[htb!]
\centering
\includegraphics[width=\columnwidth]{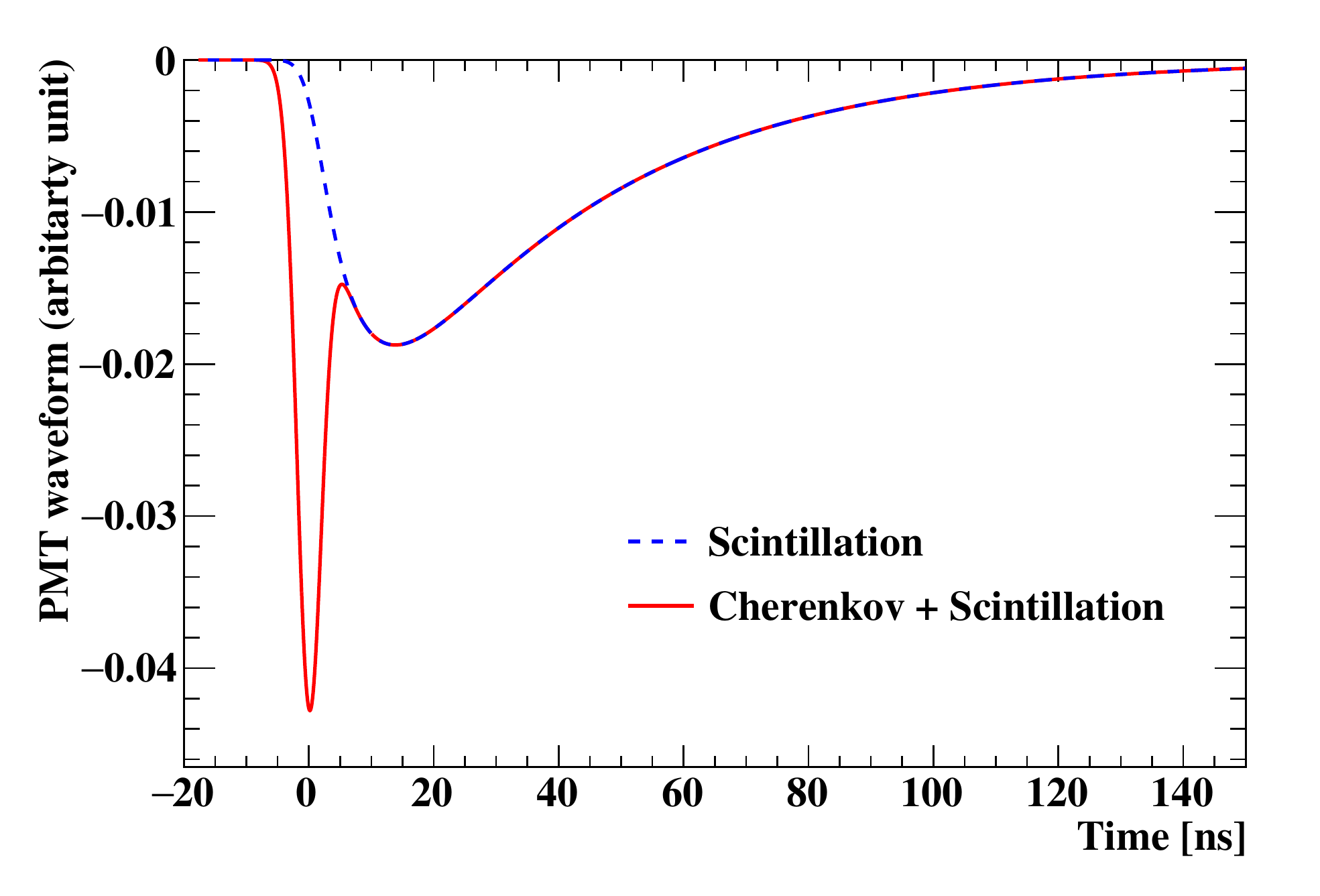}
\caption{Schematic diagram of PMT waveforms of Cherenkov and scintillation lights as shown in Ref.~\cite{LAB}. The dashed blue curve represents the PMT waveform of scintillation lights and the solid red curve represents the sum of scintillation and Cherenkov lights. The ratio of Cherenkov lights to scintillation lights is $\sim$0.2 in the case of a 10-MeV electron in the LAB.}
\label{fig:LAB}
\end{figure}

The term slow liquid scintillator refers to a type of liquid scintillator that is either water-like~\cite{WbLS, ASDC} or oil-like~\cite{LAB, LSND} and featuring a scintillator light emission time longer than its Cherenkov light emission time and the response time of common photomultiplier tubes (PMT's). Therefore, a detector with a slow liquid scintillator allows separation of scintillation light from Cherenkov light.
Ref.~\cite{LAB} demonstrates that the emission time constant of the scintillation light in LAB (illustrated in Figure~\ref{fig:LAB}) is about 37 ns, which is much longer than the Cherenkov emission time, $<$ 1 ns, and the PMT response resolution, 2 ns. The light yield of LAB is observed to be about 1000 photons/MeV.

\subsection{Particle identification}
The yields of Cherenkov and scintillation lights have different dependencies on particle types and kinetic energies.
\begin{itemize}
\item The scintillation light yield depends approximately linearly on the particle's kinetic energy; for different particles, different quenching factors should be applied according to the Birk's law~\cite{Birks}. Larger energy deposits per distance will result in heavier quenching effects.
\item The emission of Cherenkov light can be described by the formula below:
\begin{equation}
\frac{d^2N}{dx d\lambda} = \frac{2\pi\alpha z^2}{\lambda^2}\left( 1-\frac{1}{\beta^2n^2(\lambda)} \right)
\label{equ:cherenkov}
\end{equation}
where $x$ is the path length of the charged particle in the medium, $\lambda$ is the wavelength of the Cherenkov photon, $z$ is the particle charge, $\beta$ is the particle speed, $n(\lambda)$ is the index of refraction, and $\alpha$ is a constant equal to $e^2/\hbar c$. The Cherenkov light yield is directly related the speed of the charged particle in the medium and, in turn, is a function of particle's kinetic energy and rest mass.
\end{itemize}

\begin{figure*}[t]
\centering
\begin{subfigure}[h!]{\columnwidth}
	\includegraphics[width=\columnwidth]{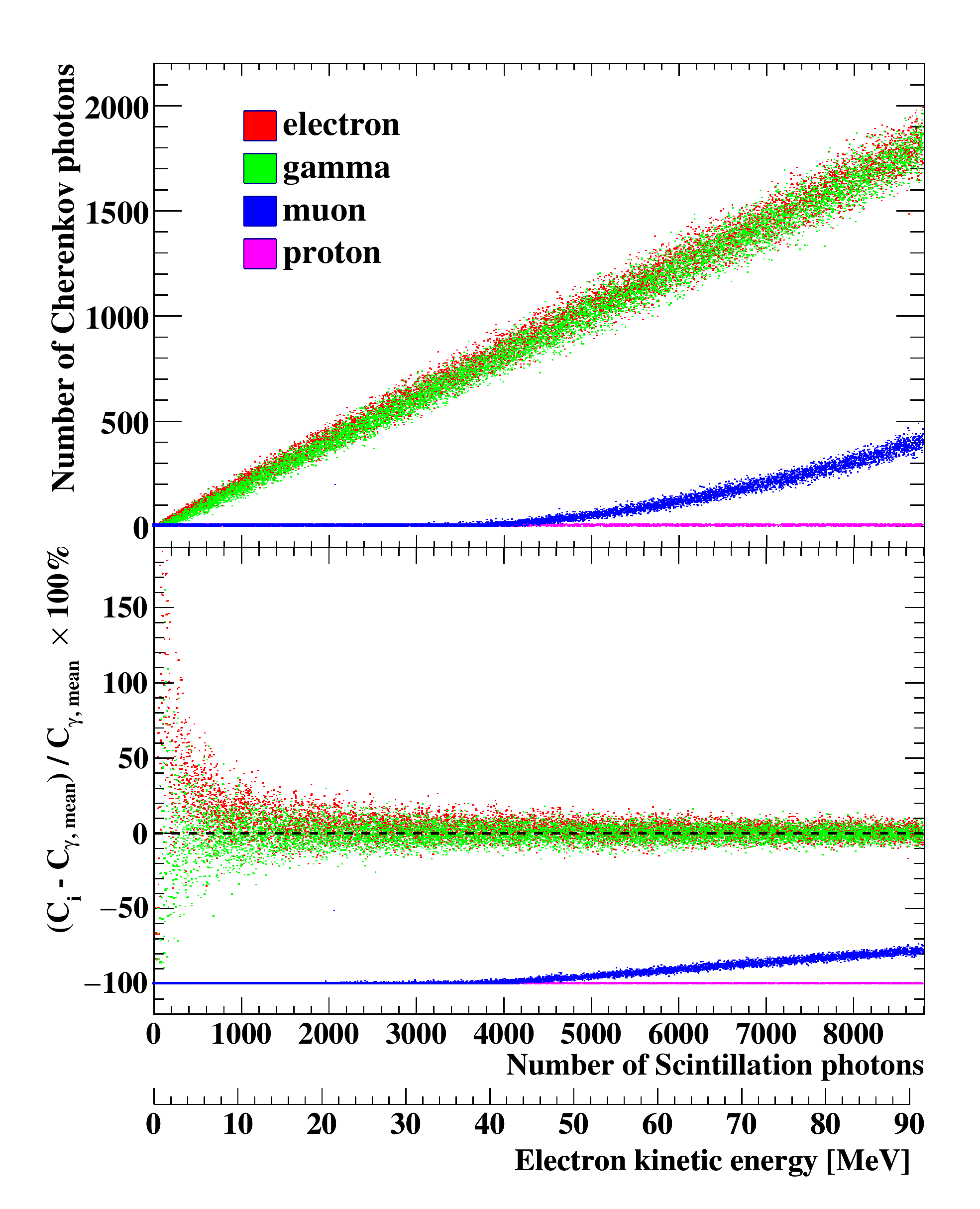}
	\caption{Ideal case}
	\label{fig:ideal}
\end{subfigure}
\begin{subfigure}[h!]{\columnwidth}
	\includegraphics[width=\columnwidth]{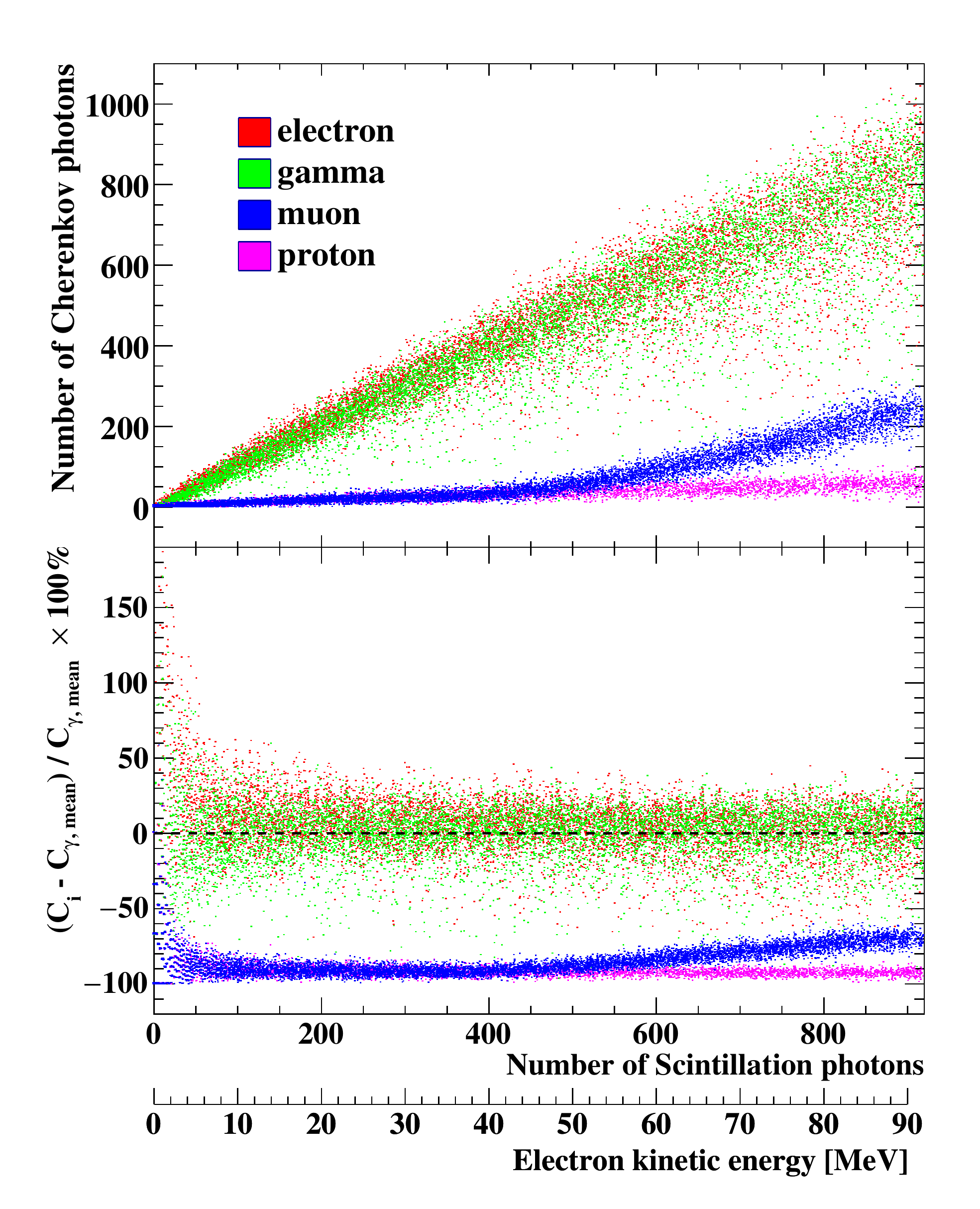}
	\caption{Realistic case}
	\label{fig:real}
\end{subfigure}
\caption{Comparison of the numbers of Cherenkov photons and the numbers of scintillation photons for various particles. A PMT detection efficiency of $\sim$10\% is assumed. (a) An ideal case reflecting the true number of accepted Cherenkov and scintillation photons. (b) A more realistic case reflecting the contamination of effect in the identification between Cherenkov and scintillation photons as well as the attenuation of optical photons in LAB according to Ref.~\cite{LABatten}. The X-axis indicates the number of scintillation photons with a scintillation light yield of $\sim$1000/MeV. The quenching effect is taken into account. Upper panel: The Y-axis indicates the number of Cherenkov photons. Lower panel: The Y-axis indicates the fractional difference of the number of Cherenkov photons to the mean value of a gamma.}
\label{fig:CS}
\end{figure*}

The concept of PID is further demonstrated with a Geant4~\cite{Geant4} simulation of
gammas, electrons, muons, and protons in a large detector filled with LAB.
The respective quenching effects (ratio of the visible energy to the deposited energy) are about 56\%, 86\%, and 96\% on average for protons, muons, and electrons in the energy of interest in the SRN study. In the upper panel of Figure~\ref{fig:CS}, we plot the number of scintillation photons versus the number of Cherenkov photons for a wavelength range of [300~nm,~500~nm], within which the PMT is assumed to have a detection efficiency of $\sim$10\%, including both the quantum efficiency and collection efficiency. 
In the lower panel of the same figure, the fractional difference of the number of Cherenkov photons to the average value of a gamma is plotted.

For a more realistic scenario, such as that in Figure~\ref{fig:real}, identification of scintillation and Cherenkov photons via the timing spectra may be affected by background contamination. A simple separation was thus conducted, i.e., photons within the first 10 ns were treated as Cherenkov photons and the rest were treated as scintillation photons. This is a conservative treatment as the pulse shape must be used for a subtle discrimination. In addition, attenuation of optical photons in LAB was taken into account assuming a kiloton-level spherical detector whose radius is at the 10-m level. According to Ref.~\cite{LABatten}, about 10\% (50\%) of the scintillation (Cherenkov) photons remains based on the scintillation emission spectrum of LAB given in Ref.~\cite{LAB} and the Cherenkov wavelength spectrum from Eq.~(\ref{equ:cherenkov}). In the lower panel of Figure~\ref{fig:real}, the up-warping structure of the muon and the proton bands for the lower energy region is due to the small ratio of the Cherenkov light to the scintillation light of an electron or a gamma.

A clear separation can be observed between electrons (gammas) and muons or protons, even for the realistic case in Figure~\ref{fig:CS}. Separation of a muon and proton can be observed in the higher energy region; here, muon has much more Cherenkov light.
It's the secondary electrons and positrons of a gamma will generate Cherenkov photons, the yield of which, however, is lower than
that of an electron with the same initial kinetic energy. This discrepancy is more significant for the case with kinetic energies less than 10 MeV and is about 1$\sigma$ significance level. Neutrons are probably accompanied by emission of gamma(s) from their scattering with a nucleus, thereby introducing Cherenkov light. The Cherenkov light hit pattern will be used to further distinguish a neutron (gamma) from an electron, as mentioned in Table~\ref{tab:key}.


\section{Sensitivity study}
\label{sec:sens}
\begin{figure}[htb!]
\centering
\includegraphics[width=\columnwidth]{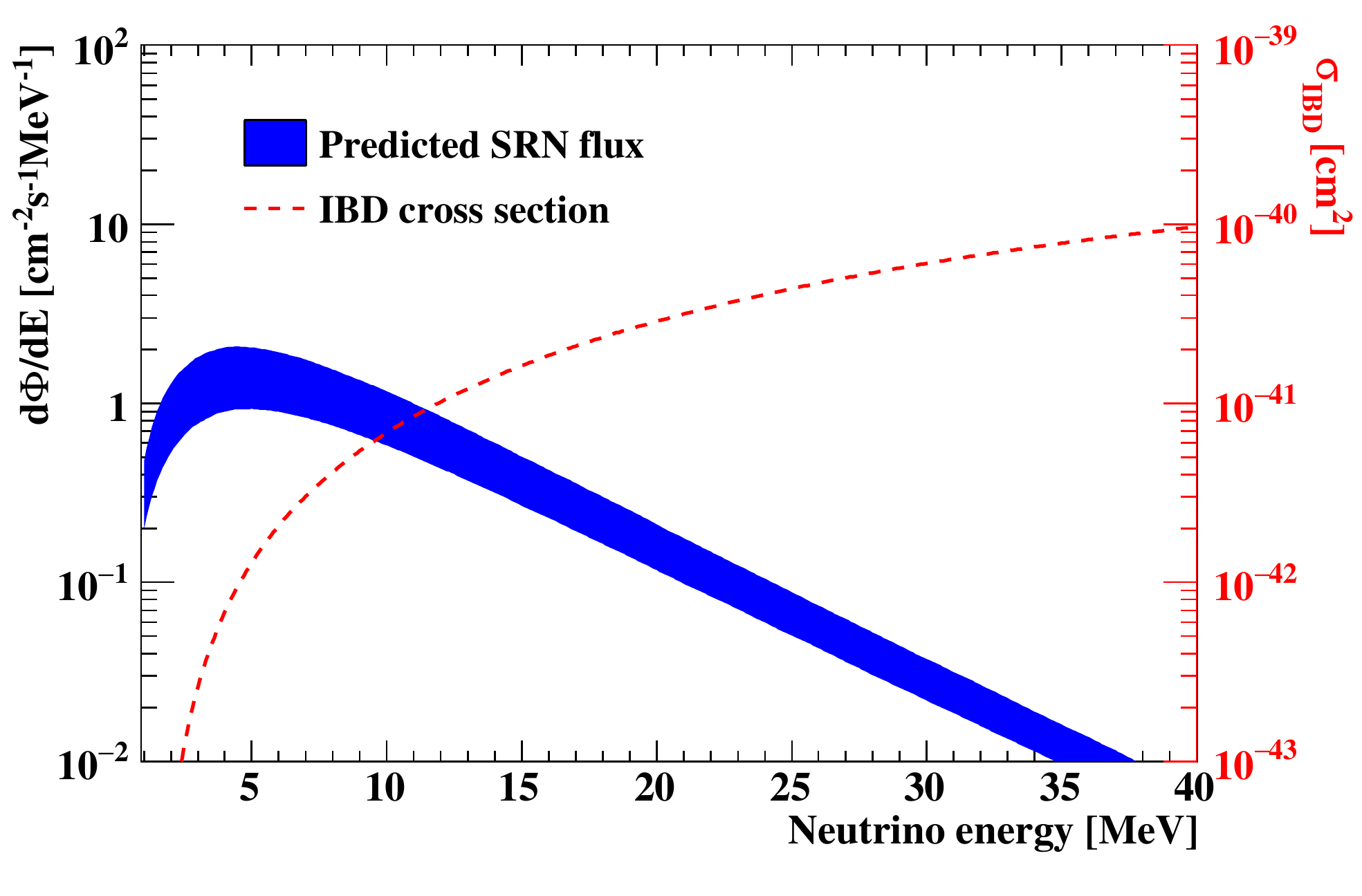}
\caption{The SRN flux predicted by the HBD (6-MeV) model with uncertainties due to astrophysical inputs~\cite{HBD}. The inverse beta decay (IBD) cross section~\cite{IBD} for anti-electron neutrino is also plotted.}
\label{fig:model}
\end{figure}

The sensitivity of the proposed SRN search is studied for a detector filled with LAB. PID is presented in Figure~\ref{fig:real}. The HBD (6 MeV) model (Figure~\ref{fig:model}) is used to predict SRN signals and demonstrate the sensitivity of our method. In this work, sensitivity is studied in the context of Jinping~\cite{Jinping, ChinaJP}, which presents two unique features: (1) it includes the deepest underground laboratory currently used and (2) it is about 1,000 km away from the nearest nuclear power plants; thus, site-independent comparisons of different detection techniques can be made.



\subsection{Signal selection and efficiency}
\label{sec:cuts}
An SRN event is identified by a prompt-delayed coincident signature from the IBD interaction. PID based on the number of Cherenkov photons (N$_{\rm Ch}$) and the scintillation photons (N$_{\rm Sc}$) is crucial to reduce the backgrounds. The selection criteria, in turn, are described below. From these criteria, 90\% selection efficiency of SRN signals can be realistically expected.

\begin{enumerate}
\item A double-coincidence within the prompt-delayed time interval of 0.2-1000~$\mu$s.
\item A cut on the N$_{\rm Sc}$ of the prompt signal, which must occur within [70, 294] counts, corresponding to the golden window of the SRN $\bar{\nu}_e$ energy (8.3-30.8 MeV $\bar{\nu}_{e}$ energy, 7.5-30.0 MeV prompt signal energy).
\item A cut on the N$_{\rm Ch}$/N$_{\rm Sc}$ of the prompt signal, which must be $>$0.65 to suppress the backgrounds with about 2\% inefficiency of the SRN events. 
\item A delayed energy cut on the gamma from neutron captures.
\item A vertex distance cut to reduce the accidental background further. 
\item The Cherenkov hit pattern (the directionality and the number of visible Cherenkov rings), mainly for the atmospheric neutrino NC background.
\end{enumerate}

\subsection{Background estimation}
\label{sec:bkg}
\subsubsection{Backgrounds from cosmic-ray muons}
A heavy overburden of $\sim$7000 m.w.e. at Jinping provides an ultra-low cosmic-ray muon flux (2$\times$10$^{-10}$/cm$^{2}$/s) with an average energy $\sim$351 GeV~\cite{Jinping}. Therefore, the spallation induced by the cosmic-ray muons can be vetoed thoroughly with a much longer muon veto window, e.g., 10-20 s, in this experiment, while negligible dead time of data taking is introduced. 

Fast neutrons are basically generated by cosmic-ray muons on the periphery of the detector and cannot be vetoed as the muon most likely does not enter the detector. However, with the ultra-low muon flux and a powerful PID in LAB, this background is estimated to be negligible for the SRN study at Jinping.

Accidental backgrounds are formed by two uncorrelated events in a detector, and can thus satisfy the IBD selection cuts in energy, time, and space. For the SRN study, the uncorrelated events in 7.5-30 MeV are mainly muon-induced fast neutrons. Therefore, the accidental backgrounds are reduced significantly and ignored at Jinping.

\subsubsection{Reactor neutrino background}
Taking into account all of the running and planned nuclear power plants currently available, the reactor neutrino flux at Jinping is estimated to be $\sim13\times10^{5}$/cm$^{2}$/s, which is quite low compared with that reported in all other experiments. However, if the energy resolution is poor, the reactor neutrino background may provide significant contributions, as shown in Figure~\ref{fig:spec}). Thus, the lower neutrino/prompt energy threshold for the SRN study must be increased to 10.8/10.0 MeV to remove the reactor neutrino background.

\subsubsection{Atmospheric neutrino backgrounds}
A simulation was performed to study the atmospheric neutrino backgrounds, convoluting GENIE-based~\cite{Genie} neutrino interaction cross sections, Geant4-based detector responses (Figure~\ref{fig:real}), and the atmospheric neutrino fluxes. The recent version 2.10.0 of Genie~\cite{GenieVersion} was used. 

The atmospheric neutrino flux was used to estimate the atmospheric neutrino CC and NC backgrounds. Above 100 MeV, the Honda flux~\cite{Honda2011} was adopted. Below 100 MeV, the flux in Ref.~\cite{Barr1988} was used despite the fact that this flux presents a large uncertainty that is highly dependent on the local environment. In fact, the contribution of the atmospheric neutrino background below 100 MeV mainly originates from $\bar{\nu}_e$'s, because the $\bar{\nu}_{\mu}/\nu_{\mu}$ CC interaction has an energy threshold of $\sim$105 MeV and neutrinos with relatively high energy could introduce the NC background because of the strong quenching effect of produced neutrons as well as that of other heavy particles in the liquid scintillator. The matter effect of neutrino oscillation~\cite{MatterEffect} was considered in the estimation of the atmospheric neutrino flux and mainly impacts the atmospheric $\nu_{\mu}$/$\bar{\nu}_{\mu}$ flux in the SRN study. 



The atmospheric $\bar{\nu}_{e}$ CC background is irreducible for the SRN study and estimated to be 0.013 event/kton-year.

With the double-coincidence cut, only $\sim$10\% of the atmospheric $\bar{\nu}_{\mu}/{\nu}_{\mu}$ CC background survives, in which case the muons do not decay within 0.2-1000 $\mu$s. From the N$_{\rm Sc}$ cut to the N$_{\rm Ch}$/N$_{\rm Sc}$ cut, less than 2\% of the events remain, in which case the Cherenkov photons are generated mainly by some Michel electrons from the muon decays within 0.2~$\mu$s. As a result, the atmospheric $\bar{\nu}_{\mu}/{\nu}_{\mu}$ CC background is about one order of magnitude smaller than the intrinsic atmospheric $\bar{\nu}_{e}$ background.

The NC interaction of the atmospheric neutrino probably produces energetic neutrons, inducing gamma(s) with adequate Cherenkov lights. With the N$_{\rm Ch}$/N$_{\rm Sc}$ cut, about 1/5 of the NC backgrounds applied with the N$_{\rm Sc}$ cut are left over. By applying the cut on the Cherenkov light hit pattern, as mentioned in Section~\ref{sec:cuts}, the above fraction can be suppressed to 3\% and the NC background is estimated to be 0.018 event/kton-year, which is roughly the same as the intrinsic atmospheric $\bar{\nu}_e$ background.

\subsubsection{Summary}
\begin{figure}[htb!]
\centering
\includegraphics[width=\columnwidth]{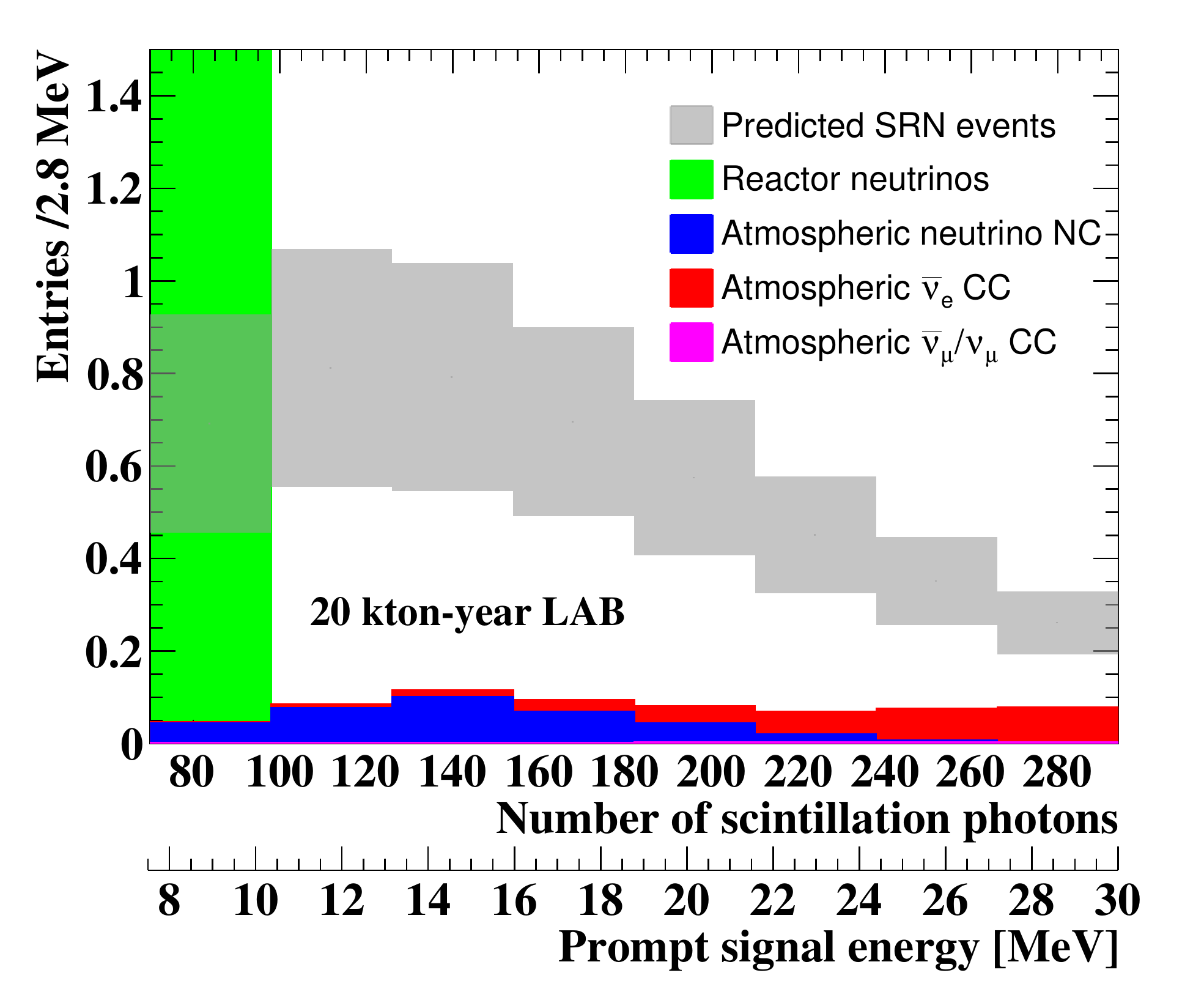}
\caption{Prompt signal energy spectra of the estimated backgrounds and the predicted SRN events in LAB with an exposure of 20 kton-year.}
\label{fig:spec}
\end{figure}

Figure~\ref{fig:spec} presents the prompt signal energy spectra of the main backgrounds and the predicted SRN events in LAB with an exposure of 20 kton-year at Jinping. The SRN events are predicted by the HBD (6-MeV) model with about 30\% uncertainty because of astrophysical inputs~\cite{HBD}. In the energy range of 10-30 MeV, the expected number of background events is about 0.1 per bin on average, in which case any event would be regarded as a ``golden'' event. Additional systematic uncertainties are not considered in this sensitivity study since the conclusion would not be essentially changed due to the very good signal-to-background ratio.

\subsection{Comparison of different techniques}
Table~\ref{tab:sum} summarizes the expected numbers of various backgrounds and SRN signals corresponding to an exposure of 20 kton-year of water, Gd-doped water, a typical liquid scintillator, and a slow liquid scintillator (LAB as a candidate) at Jinping. The neutrino energy range for the SRN study is 10.8-30.8 MeV. The backgrounds for water and Gd-doped water were estimated based on SRN analyses of the SK experiment~\cite{SKnH, SKntag, YZhangThesis}. Background estimation in KamLAND~\cite{Kam2012} was used to validate our evaluation and estimate the background for typical liquid scintillators. The signal selection efficiency, the total expected number of backgrounds, and the signal-to-background ratio are also presented. 

\begin{center}
\begin{table}[hbt!]
\centering
\begin{threeparttable}
\caption{Summary of the numbers of backgrounds and SRN events at neutrino energies of 10.8-30.8 MeV with an exposure of 20 kton-year of water, Gd-doped water, a typical liquid scintillator, and a slow liquid scintillator (LAB) at Jinping.}
\label{tab:sum}
\begin{tabular*}{\columnwidth}[]{lllll}
\hline
20 kton-year &  Water \tnote{a} & Gd-w \tnote{a} & LS & Slow LS\\
\hline
Atmos. $\bar{\nu}_{e}$ &  0.040 & 0.21 & 0.28 & 0.26 \\
Atmos. $\bar{\nu}_{\mu}$/$\nu_{\mu}$ CC &  0.33 & 1.8 & 3.6 & 0.025 \\
Atmos. NC & 0.095 & 0.49 & 62 & 0.35 \\
Total backgrounds &  0.47 & 2.5 & 66 & 0.64 \\
\hline
Signal\tnote{b} &  0.54 & 2.8 & 4.2 & 4.1 \\
Signal efficiency &  13\% & 70\% & 92\% & 90\% \\
S/B &  1.1 & 1.1 & 0.064 & 6.4 \\
\hline
\end{tabular*}
\begin{tablenotes}
\item[a] with neutron tagging.
\item[b] HBD model; water and Gd-w results corrected by a factor of $\sim$0.9 due to differences in the fractions of free protons in water and LAB.
\end{tablenotes}
\end{threeparttable}
\end{table}
\end{center}

\begin{figure}[hbt!]
\centering
\includegraphics[width=\columnwidth]{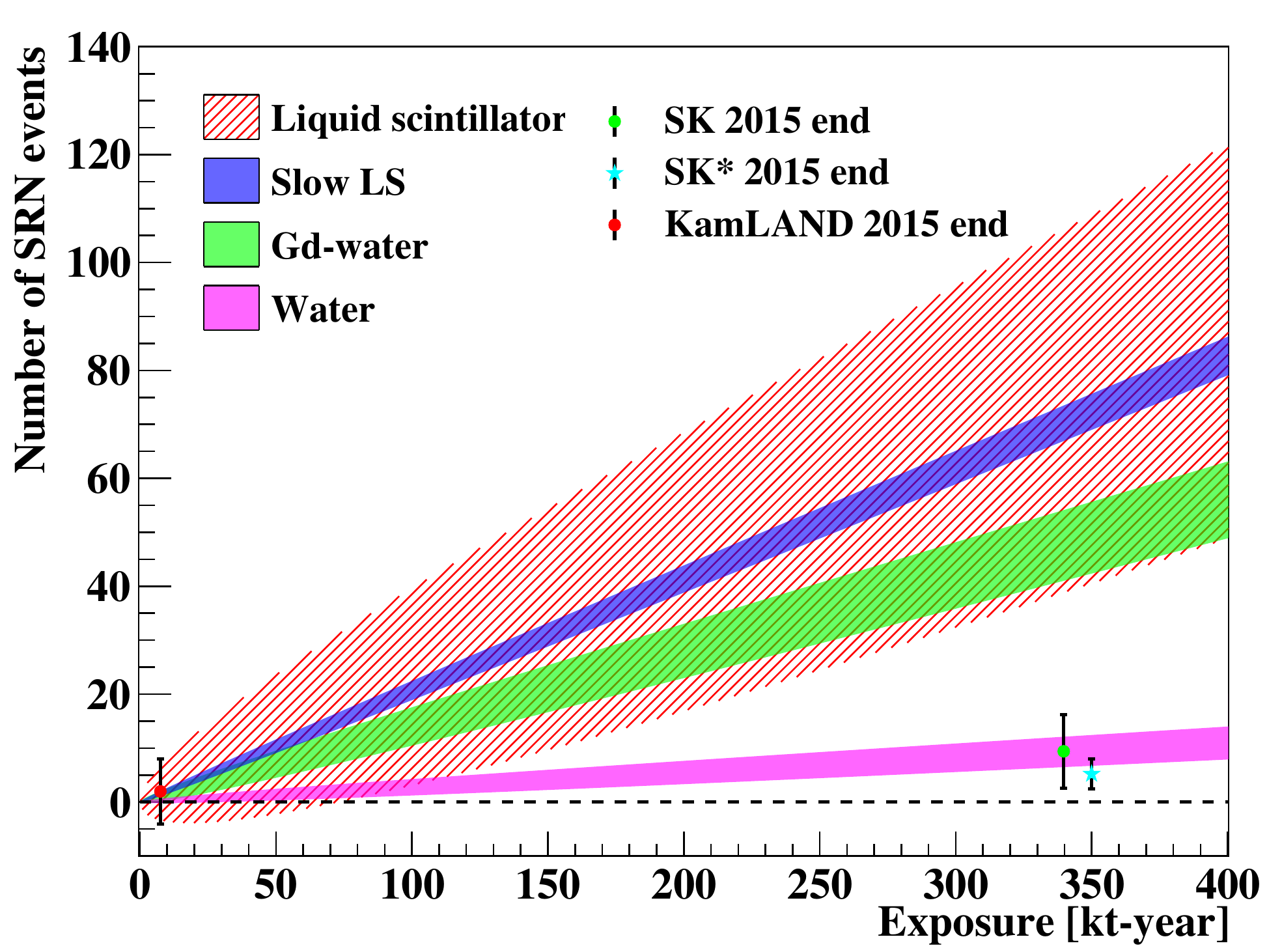}
\caption{Number of SRN signals (10.8-30.8 MeV neutrino energy) versus the exposure of several different types of detectors. The bands, whose half widths are equal to the square root of the expected numbers of total background, are drawn from the four cases presented in Table~\ref{tab:sum}. Three predicted points for KamLAND and SK experiments to the end of 2015 are also shown. The SK* point for SRN neutrino energies of 15-30 MeV is plotted by the SK point with the same exposure.}
\label{fig:result}
\end{figure}

The expected number of SRN events versus the exposure of several different types of detectors is shown in Figure~\ref{fig:result}. 
The bands, whose half widths are equal to the square root of the expected numbers of total background, are also drawn from which the significance levels of discovery are indicated. The significance levels for low-statistics cases should be cautiously calculated.
Three predicted points for KamLAND and SK experiments with data to the end of 2015 are shown according to Refs.~\cite{Kam2012} and~\cite{YZhangThesis}. The SK* point for SRN neutrino energies of 15-30 MeV is plotted by the SK point with the same exposure, reflecting the SK result excluding most of the cosmic-ray muon-induced backgrounds.

\begin{figure}[hbt!]
\centering
\includegraphics[width=\columnwidth]{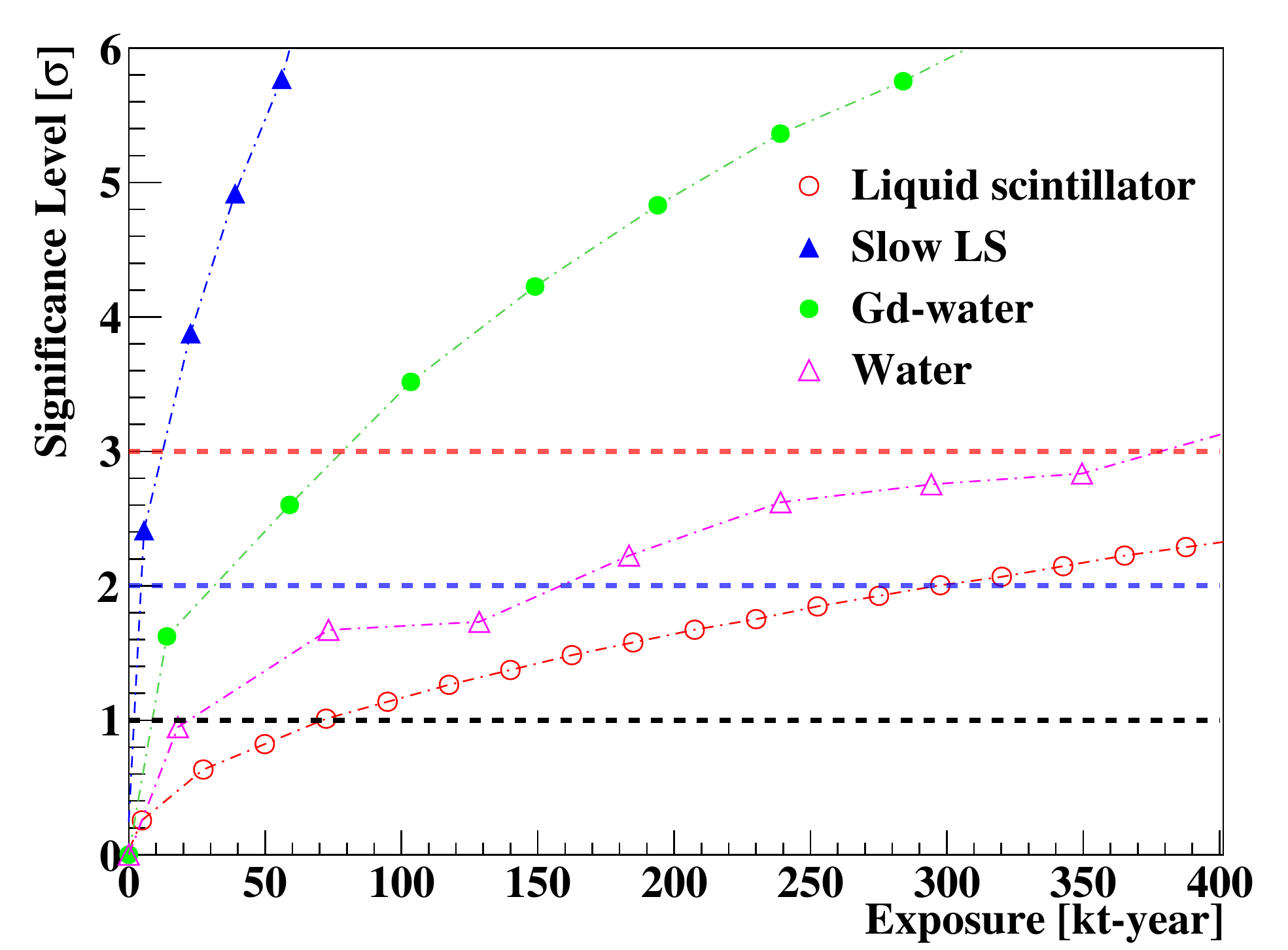}
\caption{Significance level (in units of $\chi^2$ quantile $\sigma$) versus the exposure of several different types of detectors. Three specific significance levels are indicated by the horizontal lines. There are rounding effects in the calculation of the significance levels, especially for low statistics. Certain discrete points are drawn to demonstrate the tendency for various types of detectors.}
\label{fig:sl}
\end{figure}

Against the null hypothesis, here designated as background, the significance of discovering a new signal results from the disagreement in an alternative hypothesis which includes both background as well as signal. In this paper, the significance level (P$_s$) of discovering SRNs is the total probabilities of more probable integers than the expected number of background plus signal. Assuming the expected numbers of background and SRN events are $\mu_{\rm bkg}$ and S, respectively, the definition of P$_s$ follows below,
\begin{align}
\label{eq:sl}
P_s = & \sum^{b}_{i = a}{\rm Poisson}~(i, \mu_{\rm bkg}),
\end{align}
where $i$'s are integers following Poisson distribution with the mean value of $\mu_{\rm bkg}$, S+$\mu_{\rm bkg}$ is rounded down to the nearest integer $b$, and $a$ satisfies
\begin{align}
& {\rm Poisson}(a, \mu_{\rm bkg}) \geq {\rm Poisson}(b, \mu_{\rm bkg}), \\
& {\rm Poisson}(a-1, \mu_{\rm bkg}) < {\rm Poisson}(b, \mu_{\rm bkg}).
\end{align}
For high-statistics cases, the significance levels can be easily calculated using Gaussian distributions. 
The results of the significance levels (converted to $\chi^2$ quantiles in units of $\sigma$) are shown in Figure~\ref{fig:sl}. 



In a LAB detector at Jinping, a 99.95\% (3.5$\sigma$) significance-level discovery of the SRN flux within [10.8, 30.8] MeV could be obtained with an exposure of 20 kton-years.

\section{Conclusions}
\label{sec:conclusion}
The discovery potential of supernova relic neutrinos with a slow liquid scintillator detector is discussed in this paper. Based on the separation of Cherenkov and scintillation lights by LAB, PID can be significantly enhanced to reduce the atmospheric $\bar{\nu}_{\mu}$/$\nu_{\mu}$ CC and the atmospheric NC backgrounds during SRN detection.

A kiloton-scale LAB detector at Jinping with $\mathcal{O}$(10) years of data presents excellent sensitivity for discovering SRNs. The detector shows a performance comparable with that of a large-volume Gd-doped water Cherenkov detector, as shown in Figure~\ref{fig:result}, as well as other future detectors introducing new techniques, e.g., a large-volume liquid scintillator detector with the scintillation light pulse shape discrimination to suppress the NC background~\cite{JUNOphysics, LENA-SRN} and a large-volume liquid argon time projection chamber detector~\cite{LArUS}.


\section*{Acknowledgments}
This work is supported by Key Lab of Particle \& Radiation Imaging, Ministry of Education, the CAS Center for Excellence in Particle Physics (CCEPP), Natural Science Foundation of China (nos. 11235006 and 11475093), and Ministry of Science and Technology of China (no. 2013CB834302).





\balance


\end{document}